\documentclass[fleqn,usenatbib]{mnras}

\pdfoutput=1 

\usepackage{newtxtext,newtxmath}
\usepackage[T1]{fontenc}
\usepackage{ae,aecompl}
\usepackage{hyperref} 
\usepackage{verbatim} 
\usepackage{graphicx}
\usepackage{subcaption}
\usepackage[table,usenames,dvipsnames]{xcolor}

\def\be{\begin{equation}} 
\def\ee{\end{equation}} 
\def\bea{\begin{eqnarray}} 
\def\eea{\end{eqnarray}} 

\title[Information Theory and Cosmic Strings in the CMB]{Information Theoretic Bounds on Cosmic String Detection in CMB Maps with Noise}{}

\author[R. Ciuca and O. F. Hern\'andez]{
Razvan Ciuca$^{1}$\thanks{Email: razvan.ciuca@mail.mcgill.ca}
Oscar F. Hern\'andez$^{1,2}$\thanks{Email: oscarh@physics.mcgill.ca}
\\
$^{1}$Department of Physics, McGill University, 3600 rue University, Montr\'eal, QC, H3A 2T8, Canada
\\
$^{2}$Marianopolis College,  4873 Westmount Ave.,Westmount, QC H3Y 1X9, Canada
}

\date{}
\pubyear{2019}

\begin{document}
\label{firstpage}
\pagerange{\pageref{firstpage}--\pageref{lastpage}}
\maketitle

\begin{abstract}
We use a convolutional neural network (CNN) to study cosmic string detection in cosmic microwave background (CMB) flat sky maps with Nambu-Goto strings.  On noiseless maps we can measure string tensions down to order $10^{-9}$, however when noise is included we are unable to measure string tensions below $10^{-7}$. Motivated by this impasse, we derive an information theoretic bound on the detection of the cosmic string tension $G\mu$ from CMB maps. In particular we bound the information entropy of the posterior distribution of $G\mu$ in terms of the resolution, noise level and total survey area of the CMB map. We evaluate these bounds for the ACT, SPT-3G, Simons Observatory, Cosmic Origins Explorer, and CMB-S4 experiments. These bounds cannot be saturated by any method. 
\end{abstract}


\section{Introduction}
\label{sec:intro}
Recently there has been a proliferation of methods~\citep{ Hergt:2017dr,McEwen:2017cg,VafaeiSadr:2018hh,Ciuca:2017jz,Ciuca:2019hh,Ciuca:2019ww} for the inference of the cosmic string tension $G\mu$ from 
the Gott-Kaiser-Stebbings (GKS) effect~\citep{Gott:1985eg,Kaiser:1984jg} in cosmic microwave background (CMB) temperature anisotropy maps. In~\cite{Ciuca:2017jz,Ciuca:2019hh,Ciuca:2019ww} we argued for using convolutional neural networks (CNN) to find the locations of strings in the sky and then inferred the string tension using Bayesian statistics from those estimates. We found that in noiseless maps a measurement of $G\sim 10^{-9}$ was possible. The major hurdle preventing our method and others from being applied to real instrumental data at that level of $G\mu$ sensitivity is the existence of noise in measurements of the CMB. In essentially all $G\mu$ inference methods, adding realistic noise to the simulations increases the $G\mu$ detection threshold by two or three orders of magnitude. In this paper we study cosmic string detection in CMB maps with Gaussian white noise using a CNN and derive an information theoretic bound on the $G\mu$ detection attainable in terms of the resolution, noise level and total survey area of the CMB map. 

We begin in section~\ref{sec:cosmoDetection} with a review of the recent work on the cosmological detection of cosmic strings. Readers already familiar with the subject matter may skip to section~\ref{sec:CNNnoise} where we present strategies for training convolutional neural networks in the presence of noise and present the results of string tension measurement by our network. In section~\ref{sec:infoBound} we bound the information entropy of the posterior distribution of the string tension in terms of the resolution, noise level, and area of the CMB map. With this information theory bound we can place limits on the string detection possible. We evaluate the bound and present the limits this bound places on the string tension for experiments like the 
the Atacama Cosmology Telescope (ACT)~\citep{Louis:2017il, Sherwin:2017id}, 
the South Pole Telescope 3G (SPT-3G)~\citep{Benson:2014bra},
the Simons Observatory (SO)~\citep{Ade:2019kq},
the Cosmic Origins Explorer (CORE)~\citep{Delabrouille:2018es}
and the CMB-S4~\citep{Abazajian:2019wx}. The results of these bounds are presented in table~\ref{table:experiments}.  Finally in section~\ref{sec:conclusions} we present our conclusions. 

\section{Cosmological detection of cosmic strings}
\label{sec:cosmoDetection}
Cosmic strings are linear topological defects, remnants of a high-energy phase transition in the very early Universe that  can form in a large class of extensions of the Standard Model.  Many inflationary scenarios constructed in the context of supergravity models lead to the formation of gauge theory cosmic strings at the end of the inflationary phase \citep{Jeannerot:1996ks,Jeannerot:2003er}, and in a large class of brane inflation models, inflation ends with the formation of a network of cosmic superstrings~\citep{Sarangi:2002ev} which can be stabilized as macroscopic objects in certain string models~\citep{Dvali:2004jn,Copeland:2004dw}.  
In all of the above mentioned scenarios, both a scale-invariant spectrum of adiabatic coherent perturbations and a sub-dominant contribution of cosmic strings is predicted.  Whereas cosmic strings cannot be the dominant source of the primordial fluctuations~\citep{Pen:1997fj},  they can still provide a secondary source of fluctuations. Thus searching for signatures of cosmic strings probes particle physics beyond the Standard Model in an energy range complementary to that probed by particle accelerators such as the Large Hadron Collider.  

The gravitational effects of the string can be parametrized by its string tension $G\mu$, a dimensionless constant where $G$ is Newton's gravitational constant, and $\mu$ is the energy per unit length of the string. This string tension is predicted to be between $10^{-8}<G\mu<10^{-6}$ for Grand Unified models, whereas cosmic superstrings have $10^{-12}<G\mu<10^{-6}$~\citep{Copeland:2004dw,Witten:1985tq}. 
Cosmological observations place limits on the string tension with the magnitude of the signal proportional to $G\mu$. Since the string tension is proportional to the square of the energy scale characteristic of the string, looking for cosmological signatures of cosmic strings probes particle physics with the tightest constraints on high energy physics processes.  A string moving between an observer and the surface of last scattering can lead to a step discontinuity in a CMB temperature anisotropy map through the GKS effect~\citep{Gott:1985eg,Kaiser:1984jg} of long strings. A direct search for this effect on Wilkinson Microwave Anisotropy Probe (WMAP) data led to null detection and a limit of $G\mu< 1.5 \times 10^{-6}$~\citep{Jeong:2010dh}.   More stringent  constraints come from the CMB angular power spectrum. 
The best 95\% CL constraints for the different cosmic string models are as follows. The ~\cite{PlanckCollaboration:2014il} uses a phenomenological description of Nambu-Goto strings called the Unconnected Segment Model to place an upper limit on the string tension of $G\mu < 1.3 \times 10^{-7}$, by combining its angular power spectrum measurement together with the high $\ell$ CMB information from ACT and SPT.  
For an actual Nambu-Goto string simulation, \cite{Lazanu:2015ka} has computed a limit of $G\mu < 1.5 \times 10^{-7}$.  And for Abelian-Higgs strings~\cite{Lizarraga:2016hh}
have found the limit to be $2.0\times 10^{-7}$.

The gravitational waves emitted by cosmic string loop decay provides another way to detect cosmic strings. Through pulsar timing constraints, the North American Nanohertz Observatory for Gravitational Waves (NANOgrav) placed limits on Nambu-Goto strings of $G\mu < 3.2 \times 10^{-8}$ at the 95\% CL~\citep{Arzoumanian:2016dua}. 
Such limits depend  on theoretical uncertainties such as the loop number distribution, the number of cusps and kinks per loop, and kink collisions per loop oscillation.  Reference~\citep{Ringeval:2017ew} considered these variables and the constraints that arise from both European Pulsar Timing Array (EPTA) observations and the interferometer experiment LIGO under various scenarios. Their least constrained scenario, a low kink number loop model (less than 20 kinks) with no cusps, is constrained to be $G\mu < 7.2 \times 10^{-11}$ at the 95\% CL. More stringent constraints for other scenarios are also quoted in~\citep{Ringeval:2017ew}. It should be noted that all these constraints come from Nambu-Goto string simulations where loops decay only into gravitational waves. Abelian Higgs strings evade these constraints since they decay by particle emission~\citep{Hindmarsh:2018gi}.
For this reasons, and the continued theoretical uncertainty in cosmic string loop variables, limits from cosmic string loops are considered less robust than those arising from long strings, which is why the robust limit is still that provided by the Planck collaboration $G\mu\lesssim10^{-7}$. 

Much research has been done to find a more sensitive probe of cosmic strings in CMB and 21 cm intensity maps.  As long cosmic strings move, they accrete matter into over-dense wakes which perturb the CMB light and the 21 cm line in particular. Future 21 cm redshift surveys could observe cosmic strings wakes through their distinctive shape in redshift space~\citep{Brandenberger:2010hi,Hernandez:2011ima,Hernandez:2012gz,Hernandez:2014cu,daCunha:2016bo} or through the cross-correlation between CMB and 21 cm radiation from dark ages~\citep{Berndsen:2010ku}.  Edge and shape detection algorithms such as the Canny algorithm~\citep{Canny:et}, wavelets, and curvelets have been proposed and studied as alternatives to the power spectrum in looking for cosmic strings in these maps~\citep{Amsel:2008it,Stewart:2009fr, Hergt:2017dr,McEwen:2017cg,VafaeiSadr:2018hh}. 

Finally machine learning has been applied to the search for cosmic strings in CMB temperature maps~\citep{Ciuca:2017jz, Ciuca:2019ww,VafaeiSadr:2018bc,Ciuca:2019hh}. These are the techniques that have achieved the best results so far. The authors of \cite{VafaeiSadr:2018bc} used tree-based machine learning algorithms to place measurement limits of $1.2\times10^{-7}$ and $3.6\times10^{-9}$ for 0.9 arcmin resolution maps with noise and without noise, respectively. 
\footnote{\cite{VafaeiSadr:2018bc} also quote detection limits of $3.0\times10^{-8}$ and $2.1\times10^{-10}$ in maps with and without noise, respectively. The detection limit differs from their measurement limits (see section 2.3 and their tables 1 and 2). It is their measurement limit that is most closely related to the string tension limits we quote in our work.}
The approach we took in ~\cite{Ciuca:2017jz,Ciuca:2019hh, Ciuca:2019ww} was to develop and train a convolutional neural networks to estimate the locations of strings in a sky map. From the CNN estimates of string locations we inferred the string tension using Bayesian statistics. In ~\cite{Ciuca:2019hh} we achieved a $G\mu$ measurement limit of $2\times10^{-9}$ in noiseless 1 arcmin maps.  We now turn to the next section where we describe our results for string detection in CMB maps with noise using our CNN. We will show that with 1.5~$\mu$K~arcmin Gaussian white noise we can measure the string tension of strings with $G\mu$ as low as $10^{-7}$. This is numerically the same measurement as in \cite{VafaeiSadr:2018bc} but for an experiment with smaller survey area and greater noise. 

\section{A CNN for string detection in noisy CMB maps}
\label{sec:CNNnoise}
The string temperature map used in our previous work~\citep{Ciuca:2017jz,Ciuca:2019hh,Ciuca:2019ww} was based on a simulation with long straight strings. Here we use the same CNN but work with Nambu-Goto simulations for string temperature maps which were provided to us by Fran\c{c}ois Bouchet and Christophe Ringeval~\citep{Ringeval:2012gp,Fraisse:2007nu}. We model the CMB flat sky map  $\delta_{sky}$ as being composed of two different elements, $\delta_{gauss}$ and $\delta_{str}$. The $\delta_{gauss}$ term is the standard $\Lambda$CDM cosmology CMB anisotropies that can be computed from the power spectrum, whereas $\delta_{str}$ is made up of the superposition of GKS temperature discontinuties of individual, strings as described in \cite{Fraisse:2007nu}. These are $1024\times1024$ pixel maps with resolution of 0.420 arcmin, and thus represent a survey area of (7.168 deg)$^2$. We will call these flat sky maps Ringeval-Bouchet maps. 
Finally we add a Gaussian white noise component $\sigma\, \delta_{noise}$, so that
\be
\label{eq:stringmodel}
\delta_{sky} = \delta_{gauss} + G\mu\, \delta_{str} + \sigma\, \delta_{noise} \, .
\ee

We again find that in noiseless maps we can measure the string tension of strings with $G\mu\gtrsim 10^{-9}$. However noise rapidly deteriorate the accuracy of string location estimates and therefore the estimates of the string tension. To obtain better results we tried various training strategies, which we will describe below. 
With the best of these approaches, we were able to measure string tensions as low as $G\mu\sim10^{-7}$ for maps with noise comparable to that in the data from the South Pole Telescope 3G (SPT-3G)~\cite{Benson:2014bra}.
The fact that the string tension limits obtained with noise are so much weaker than the noiseless case, led us to question what limits information theory is setting on the cosmic string tensions detectable in CMB maps, given the size, resolution and noise level per pixel. The exploration of this question is presented in section~\ref{sec:infoBound}. 
Here we describe how we adapted the residual neural network described in~\cite{Ciuca:2019hh} to detect strings in CMB maps with noise.

In extending our neural network to maps with noise, getting the training process to converge became significantly more difficult. 
To improve the convergence of the training, we tried various heuristics: 
\begin{itemize}
\item start with a network trained on noiseless maps and increase the noise gently,
\item vary the learning rate schedule in various ways,
\item force the output of the network to have a fixed variance across the map,
\item start  training with an extremely large $G\mu$ and decrease this slowly as we proceed in the training.
\end{itemize}
It was this last factor that proved crucial for convergence. We began the training so that the contribution of strings to the signal would be dominant over both the Gaussian fluctuations and the noise. In practice we accomplished this by starting with $G\mu = 5\times 10^{-6}$.  Hence at the beginning of training the network learns to pick out strings from the pure string temperature map, without the added difficulty of needing to filter out the noise. Once the network has learned to do that much, $G\mu$ is decreased and thus the relative contribution of the noise and the Gaussian fluctuations increases.We decreased $G\mu$ every $2000$ training iterations. 

We should note that when we started the training at too low a $G\mu$, all the network  learned was what fraction of all the pixels had strings, and it would predict that fraction as the probability that any pixel would have a string, for all pixels. This held true no matter the network architecture. When we trained on maps without noise, we needed to start our training at $G\mu= 10^{-7}$ to avoid this effect.  

The structure of the CNN we used with noise is the same as in \cite{Ciuca:2019hh}. We take the basic structure of the network to be a sequence of residual blocks, each of whose main part is an $n\times n$ convolution or filter.
With low noise, the easiest way to detect strings is to exclusively look at temperature discontinuities on small scales. In this regime, a small neighbourhood of the string is required to determine that it is in fact a string. As we increase the noise, this method will stop working. Looking at a small neighbourhood around a string means that we do not have enough data to overpower the noise. 

One idea is to try to take advantage of the specific large-scale structure of the string signal by looking at a much larger neighbourhood of a particular pixel to determine whether it is on a string. The maximum spatial scale on which the network can make inferences is completely determined by the depth of the network and the filter size used within each convolution. A deeper network with wider filters is better equipped to learn about this large scale structure because they are able to do computations with larger spatial correlations. We experimented with increasing the filter width and the number of layers in the network to allow the convolutional neural network to look at very large neighbourhoods in order to  predict string locations. Training these networks was computationally costly and yielded no performance benefits. We found no difference between a 10-layer network and a 30-layer network. In the end we took the $n$ of the filter to be $3$ as we did in the noiseless case. 
  
\begin{figure}
  \includegraphics[width=\linewidth]{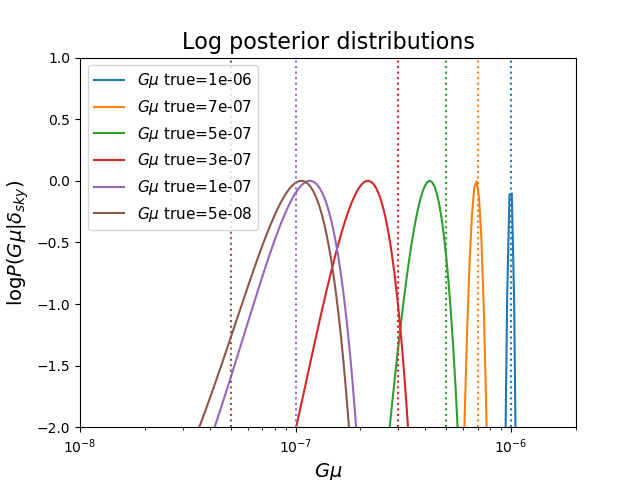}
  \caption{Posteriors produced by the 10-layer network on a survey area of  $A=514$~deg$^2$ with a resolution of $a=$~1.0 arcmin and noise $\sigma=1.5$~$\mu$K arcmin. }
  \label{fig:noisePosteriors}
\end{figure}

To compute the posterior distribution for a simulated CMB sky map with survey area $A$, resolution $a$, and noise $\sigma$ we proceed as follows. We convolve the Ringeval-Bouchet map (equation~\eqref{eq:stringmodel}) with a Gaussian beam whose FWHM$=a$. We use as many Ringeval-Bouchet maps as necessary to cover the survey area and for each map we have the CNN compute the log posteriors. Each of these maps has a different $\Lambda$CDM, string, and noise component. For each log posterior, we find its maximum and fit a quadratic to a neighbourhood around it.  The average and variance of these maxima give us the parameters of a Gaussian distribution, which is what we take to be the predicted log posterior.  In figure~\ref{fig:noisePosteriors} we show the posterior probabilities produced by the network on 10 Ringeval-Bouchet maps with $1.5 \mu$K armin of noise convolved to 1.0 arcmin resolution.  This corresponds to a survey area of  514 deg$^2$. The SPT-3G experiment has approximately the same resolution and noise but 5 time the survey area. 
%
We plot this for six different values of the true string tension in the map.  
For $G\mu=10^{-7}$ the string tension is still measurable,
\be
\label{eq:gmu1e7}
\langle G\mu \rangle = \int G\mu \, P(G\mu | \delta_{sky})~ d(G\mu)= 1.17 \times10^{-7} \ ,
\ee
since the posterior probability $P(G\mu | \delta_{sky})$ is a Gaussian with mean $1.17\times10^{-7}$ and standard deviation of $0.37\times10^{-7}$.

\section{An information theory bound on the string tension in a CMB map}
\label{sec:infoBound}
The fact that the string tension limits obtained with noise are so much weaker than the noiseless case suggests we should investigate the limits information theory would set on the cosmic string tensions detectable in CMB maps. 
To make the calculation feasible, we will calculate the information entropy of a map assuming perfect knowledge of the GKS induced string temperature map. 
Conditioning on the temperature contribution of strings amounts to assuming that we have knowledge of all aspects of how strings influence the CMB, except for $G\mu$. 
Yet even with such a drastic assumption it turns out that it is impossible to detect smaller values of $G\mu$ than about $10^{-10}$ in the data from the SO or the SPT-3G. Since perfect knowledge of the string temperature map contribution is unattainable, the actual detection limits we can expect would be much weaker. 

The information entropy $H(X)$ of a discrete random variable $X$ is given by
\be
\label{eq:entropydef}
H(X) \equiv - \sum_{x} p(x)\log p(x) \ . 
\ee
This quantity is always positive and is equal to zero if and only if $p(x)$ is a delta function. We can extend this entropy definition to continuous random variables by simply turning the sum into an integral sign, but we lose the positivity guarantee in doing this. We define the concept of conditional entropy $H(X|Y)$ by
\begin{multline}
H(X|Y) = \sum_{y}p(Y=y)H(X|Y=y) \\
= - \sum_{y, x} p(y)p(x|y) \log p(x|y) =  - \sum_{y, x} p(x,y) \log p(x|y)
\end{multline}
It is easy to prove that $H(X|Y) \leq H(X)$ and that $H(X|Y,Z) \leq H(X|Z)$, where $X$, $Y$ and $Z$ are any set of random variables. This simply states the fact that on average, knowing the value of $Y$ cannot diminish our knowledge of $X$.

Using this conditional entropy inequality and applying it to the observation of $G\mu$, we can obtain a simple bound on the minimum entropy of $P(G\mu | \delta_{sky})$. While computing the posterior on $G\mu$ given the observed CMB sky map requires difficult statistics, computing the same posterior becomes easier if we also condition on the temperature contribution of strings. Since the entropy of the latter is always smaller than that of the former, we can constrain properties of the former difficult posterior. 
Applying the conditional entropy bound to the string tension $G\mu$, the total CMB temperature $\delta_{sky}$ and the $G\mu$ independant string contribution $\delta_{str}$, we get
\be
\label{eq:simplebound}
H(G\mu ~ | ~ \delta_{sky}) \geq H(G\mu ~ | ~ \delta_{sky}, \delta_{str})
\ee
Given samples $G\mu$, $\delta_{gauss}$ , $\delta_{str}$, $\delta_{noise}$ from $P(G\mu)$, $P(\delta_{gauss})$, $P(\delta_{str})$, $P(\delta_{noise})$, respectively. 
we can generate a sample of $\delta_{sky}$ by combining these maps with equation~\eqref{eq:stringmodel}.
Given that generating $\delta_{sky}$ involves  multiplying $\delta_{str}$ by $G\mu$, conditioning on $\delta_{str}$ in the posterior $G\mu$ computation does not completely determine the string temperature contribution to the sky map since we are still missing the overall scale of the string contributions, given by $G\mu$. The general idea of the bound is to constrain the range of $G\mu$ with CMB observations, noting that we can never constrain $G\mu$ better than if we had access to $\delta_{str}$. 

Next, by the definition of conditional entropy we have
\begin{multline}
H(G\mu ~ | ~ \delta_{sky}, \delta_{str}) = 
\sum_{x,y} P(\delta_{sky}=x, \delta_{str}=y)
\\ \times H(G\mu|\delta_{sky}=x, \delta_{str}=y)\ .
\end{multline}
We take the data set of Ringeval-Bouchet maps to be a representative sample drawn from $P(\delta_{sky}=x, \delta_{str}=y)$, and thus
\be
\label{eq:conditionalentropy1}
H(G\mu ~ | ~ \delta_{sky}, \delta_{str}) = 
\sum_{x,y \in dataset} H(G\mu|\delta_{sky}=x, \delta_{str}=y)\ ,
\ee
where
\begin{multline}
\label{eq:conditionalentropy2}
H(G\mu|\delta_{sky}=x, \delta_{str}=y) 
 =  -\sum_{G\mu} P(G\mu | \delta_{sky}=x, \delta_{str}=y)
\\ \times \log P(G\mu | \delta_{sky}=x, \delta_{str}=y)\ . 
\end{multline}
The quantity $P(G\mu | \delta_{sky}, \delta_{str})$ is given by Bayes' rule:
\be
P(G\mu ~ | ~ \delta_{sky}, \delta_{str}) \propto P(\delta_{sky} ~ | ~ G\mu, \delta_{str}) P(G\mu | \delta_{str})\ . 
\ee
The value of $G\mu$ does not depend on the string map,  hence $P(G\mu | \delta_{str})$ is  constant, and we have
\be
P(G\mu ~ | ~ \delta_{sky}, \delta_{str}) \propto P(\delta_{sky} ~ | ~ G\mu, \delta_{str}) \ . 
\ee

We know that $\delta_{noise}$ is simply pixel-independent Gaussian noise, and that $P(\delta_{gauss})$ is also a product of independent normal distributions if we go to Fourier space. The only quantity with a distribution which is not known exactly is $\delta_{str}$, we are bypassing that restriction by simply conditioning on $\delta_{str}$. Letting $\tilde{\delta}_{str}$ be the Fourier transform of the string map, we have the following:
\begin{multline}
\log P(\delta_{sky} ~ | ~ G\mu, \delta_{str})
\\  
= -\sum_{k_x, k_y} \frac{\bigg(\tilde{\delta}_{sky}(k_x, k_y) - G\mu\, \tilde{\delta}_{str}(k_x, k_y) \bigg)^2}{2 (C_{g}(k_x, k_y) +
\sigma^2)} 
\\ 
- \sum_{k_x, k_y} \frac{1}{2} \log (2\pi (C_{g}(k_x, k_y) + \sigma^2) )
\end{multline}

That is, the probability of the $(k_x, k_y)$ mode is simply Gaussian with a mean equal to the known string contribution and a variance given by the power spectrum and the noise. Note that the second term is redundant, since it is $G\mu$-independent and we will need to normalise the distribution over $G\mu$. We can thus neglect it. Note also that the variance of the Gaussian modes only depend on $k=\sqrt(k^2_x+k^2_y)$.
\begin{multline}
\label{eq:range}
\log P(\delta_{sky} ~ | ~ G\mu, \delta_{str}) 
\\
= -  \sum_{k_x, k_y}
\frac{\bigg(\tilde{\delta}_{sky}(k_x, k_y) - G\mu\, \tilde{\delta}_{str}(k_x, k_y) \bigg)^2}{2 (C_{g}(k) +
\sigma^2)} 
\end{multline}
%
Writing $G\mu^*$ for the real $G\mu$ of the sky map we are observing:
\begin{multline}
\log P(\delta_{sky} ~ | ~ G\mu, \delta_{str}) 
\\ = -  \sum_{k_x, k_y} 
\frac{\big[\tilde{\delta}_{g+n}(k_x, k_y) + (G\mu^* - G\mu)\tilde{\delta}_{str}(k_x, k_y) \big]^2}{2 (C_{g}(k) +
\sigma^2)} 
\end{multline}
Where $\delta_{g+n}$ is just the Gaussian fluctuations plus the white noise. This is uncorrelated with $\delta_{str}$, therefore
\begin{multline}
\log P(\delta_{sky} ~ | ~ G\mu, \delta_{str}) 
\\ =-  \sum_{k_x, k_y} 
\frac{\big[\tilde{\delta}_{g+n}(k_x, k_y)\big]^2 + \big[(G\mu^* - G\mu)\tilde{\delta}_{str}(k_x, k_y) \big]^2}{2 (C_{g}(k) +
\sigma^2)}
\end{multline}
but we only care about $G\mu$-dependent terms, so we can ignore the terms added by $\delta_{g+n}$, giving us
\begin{multline}
\log P(\delta_{sky} ~ | ~ G\mu, \delta_{str}) 
\\
=-(G\mu^* - G\mu)^2   \sum_{k_x, k_y}
\frac{ \big[\tilde{\delta}_{str}(k_x, k_y)\big]^2}{2 (C_{g}(k) +
\sigma^2)}
\end{multline}

Given that $G\mu$ cannot be negative, we have that the probability distribution for $G\mu$ conditioned on the sky and string map is a truncated normal distribution centred around the correct $G\mu$ with a variance that depends on the the power spectrum of the string map. Conditioning on $\delta_{str}$ has allowed us to simplify what would have been a nightmarish computation into a closed form depending only on the power spectra of Gaussian fluctuations and string contributions. To obtain $H(G\mu ~ | ~ \delta_{sky}, \delta_{str})$ we use equations~\eqref{eq:conditionalentropy1} and \eqref{eq:conditionalentropy2} to compute the average of the entropy of this posterior over $G\mu^*$, $\delta_{sky}$ and $\delta_{str}$. 
Since the summand doesn't involve the explicit Gaussian and noise maps, $P(G\mu | \delta_{sky}=x, \delta_{str}=y)$ marginalises to $P(G\mu | \delta_{str}=y)$. We thus have 
\begin{multline}
\label{eq:boundtruncated}
H(G\mu|\delta_{sky}, \delta_{str}) 
\\ =  \sum_{ \substack{ \scriptscriptstyle G\mu^* \\ \scriptscriptstyle \delta_{str} \in dataset} } 
\exp{\Big(-(G\mu^* - G\mu)^2   \sum_{k_x, k_y}
\frac{ \big[\tilde{\delta}_{str}(k_x, k_y)\big]^2}{2 (C_{g}(k) +
\sigma^2)}\Big)}
\\ \times \Bigg((G\mu^* - G\mu)^2   \sum_{k_x, k_y}
\frac{ \big[\tilde{\delta}_{str}(k_x, k_y)\big]^2}{2 (C_{g}(k) +
\sigma^2)}\Bigg)\ . 
\end{multline}
Because of the truncated Gaussian, we will evaluate this entropy numerically. However if we ignore the fact that our posterior is truncated, we can use the fact that the entropy of a Gaussian distribution of variance $\sigma^2$ is $\log \sigma (\sqrt{2\pi e})$ to approximate the entropy as
\begin{multline}
H(G\mu|\delta_{sky}, \delta_{str}) 
\\ \approx 
\sum_{  \scriptscriptstyle \delta_{str} \in dataset} 
-\frac{1}{2}\log \Bigg (\frac{1}{2\pi e} 
 \sum_{k_x, k_y} 
\frac{ \big[\tilde{\delta}_{str}(k_x, k_y)\big]^2}{2 (C_{g}(k) +
\sigma^2)}
\Bigg) \ .
\end{multline}
Combining this with equation~\eqref{eq:simplebound}, we get the bound
\begin{multline}
\label{eq:bound}
H(G\mu ~ | ~ \delta_{sky}) 
\\ \gtrsim 
\sum_{  \scriptscriptstyle \delta_{str} \in dataset} 
-\frac{1}{2}\log \Bigg (\frac{1}{2\pi e} 
   \sum_{k_x, k_y} 
\frac{ \big[\tilde{\delta}_{str}(k_x, k_y)\big]^2}{2 (C_{g}(k) +
\sigma^2)} \Bigg) \ .
\end{multline}

To evaluate the entropy bound numerically we use a data set of 100 Ringeval-Bouchet string temperature maps. These maps have $1024\times 1024$ pixels with a resolution of $a=0.42$~arcmin and an area $A=51.4$~deg$^2$. We convolve them with a Gaussian beam whose FWHM = 1 arcmin, to produce $512\times 512$ pixel maps with resolution of 1 arcmin and with the same area as before. We then take the Fourier transform of these maps numerically. We evaluate this entropy bound for each of the 100 maps in our data set, and then take the average the entropy values as given in equation~\eqref{eq:boundtruncated}.

To obtain an estimate of this Fourier transform for a string temperature map of a sky area $A'$ greater than $A$ we could use more string maps from our data set. However we need to use the limited number of maps in our data set to take the average of the entropies we calculated numerically. Hence we approximate the increase in area by scaling $\tilde{\delta}_{str}(k_x, k_y) \to \sqrt{A'/A} ~\tilde{\delta}_{str}(k_x, k_y)$.  Thus the entropy will scale by $\log(A'/A)$.
The argument that justifies this approximation is as follows. The string signal is due to the short distance GKS effect and there is no information in the long scale correlations. Thus we expect the small wavenumber modes to give contribute very little to the Fourier sum. For the high wave number modes that do contribute, we wish to approximate the contribution of 2 independent modes, $\delta_k$ and $\delta'_k$, by $2\times\delta_k$. 
Since each mode is sampled from a Gaussian distribution with the same mean and standard deviation $\sigma_k$. The error in this approximation, $\delta'_k-\delta_k$, is also a Gaussian with mean zero and standard deviation $\sqrt{2} \sigma_k$. Thus by simply scaling from area $A$ to $A'$ the error has a standard deviation of $\sqrt{A'/A}~\sigma_k$ and this is the error we have in approximating the independent modes by $\sqrt{A'/A}~\delta_k$. Since $\sigma_k\sim \delta^2_k$ and $\delta_k<1$, this error is small. 

To interpret what this bound implies for string detection possibilities, consider a statistical method for obtaining the posterior of $G\mu$ given the observed CMB. For simplicity, imagine that this method only produces uniform posterior distributions on the interval $[0, G\mu']$. The entropy of a uniform posterior is simply $\log G\mu'$. Since no method of statistical inference for the string tension from the CMB can produce posteriors of width smaller than that allowed by the bound, $\log G\mu'$ cannot be smaller than the right hand side of equation~\eqref{eq:boundtruncated}. We can associate a given information entropy value $H$ to a $G\mu$ detection bound by noting that a uniform distribution on $[0, \exp H]$ has entropy $H$. So a bound on the entropy of the $G\mu$ posterior can roughly be associated to a $G\mu$ limit of $\exp H$. The results of this bound applied to different experiment can be found in table~\ref{table:experiments}.

As we stated before, this bound is produced assuming 
perfect knowledge of $\delta_{str}$. This extra knowledge is so strong that it is likely that the best $G\mu$ inference method will be much weaker than the bound. 
By the chain rule, we can decompose the computation of $P(G\mu | \delta_{sky})$ into 
\be
P(G\mu | \delta_{sky}) = \sum_{\delta_{str}} P(G\mu | \delta_{sky}, \delta_{str})P(\delta_{str} | \delta_{sky})
\ee
Computing $P(G\mu | \delta_{sky}, \delta_{str})$ does not involve any difficulty and follows exactly from \eqref{eq:stringmodel}. The full difficulty resides in computing $P(\delta_{str} | \delta_{sky})$. The bound is derived by assuming that $P(\delta_{str} | \delta_{sky})$ is a delta function around the correct string map. 

\begin{table}{
 \begin{tabular}
{ p{1.25cm}||p{0.8cm}|p{1.15cm}|p{1.1cm}|p{1.0cm}| p{1.23cm}  } 
 \hline
 Experiment & Survey Area [$\text{deg}^2$] & Resolution [arcmin] & Noise [$\mu K$ arcmin ] & Entropy Bound [nats] & Implied $G\mu$ bound \\ [0.5ex] 
 \hline\hline 
 ACT D6 &~~~71    & ~~~1 & 10 &  $-19.79$   &  $ 2.5\times 10^{-9}$    \\
 ACT D56 &~~~626    & ~~~1 & 17 &  $-20.49$   &  $ 1.3\times 10^{-9}$    \\
 SO  & 16501   & ~~~1 & ~~6  & $-22.98$ & $ 1.0\times 10^{-10}$ \\ 
 SPT-3G &~~2500& ~~~1  & ~~1.6 & $-23.15$  & $ 8.9 \times 10^{-11}$ \\ 
 CORE & 20627 &  ~~~2 & ~~2 & $-23.85$  &  $4.4\times 10^{-11}$ \\ 
 CMB-S4 & 24752 & ~~~1 & ~~1  & $-24.69 $ & $1.9\times 10^{-11}$ \\ [1ex] 
 \hline
\end{tabular}
 \caption{Bounds for 
the Atacama Cosmology Telescope (ACT)~\citep{Louis:2017il, Sherwin:2017id}, 
the South Pole Telescope 3G (SPT-3G)~\citep{Benson:2014bra},
the Simons Observatory (SO)~\citep{Ade:2019kq},
the Cosmic Origins Explorer (CORE)~\citep{Delabrouille:2018es}, 
and the CMB-S4~\citep{Abazajian:2019wx}.}  
\label{table:experiments}
  }
\end{table}

\section{Conclusions}
\label{sec:conclusions}

In this paper we described the string detection results obtained with the CNN we developed in~\cite{Ciuca:2017jz,Ciuca:2019hh,Ciuca:2019ww} when applied to CMB maps with Gaussian white noise. We have trained and evaluated our CNN using Nambu-Goto simulations \citep{Ringeval:2012gp,Fraisse:2007nu} instead of the straight string simulations of our previous work. 
The best string tension measurement result in CMB maps with noise that we have found in the literature is $1.2\times10^{-7}$ for a simulation corresponding to a CMB-S4 experiment~\cite{VafaeiSadr:2018bc}. We are able to match this result for an experiment with a smaller survey area and higher noise. In particular we found that an experiment such as the SPT-3G should be able to measure strings with tensions greater than or equal to $1\times10^{-7}$ (see figure~\ref{fig:noisePosteriors} and equation~\eqref{eq:gmu1e7}). 

Since string tensions below order $10^{-7}$ are already excluded by the Planck power spectrum constraints~\citep{PlanckCollaboration:2014il}, the results quoted above do not really improve detection limits.  This motivated our search for an information theory limit on the string detection possible in CMB maps with noise. 
We defined the performance of an inference method as the information entropy of the posterior probability $P(G\mu | \delta_{sky})$ produced by the method, equation~\eqref{eq:entropydef}. We then derived a bound on this entropy and consequently a detection bound on the string tension, equation~\eqref{eq:bound}.
The results of this bound when applied to different experiments is summarized in table~\ref{table:experiments}.  
The Standard Model extensions in~\cite{Copeland:2004dw,Witten:1985tq} that produce cosmic strings have a string tension lower bound of $G\mu > 10^{-12}$. None of the CMB experiments can reach this bound. ACT gives bounds of the order of $10^{-9}$, SO and SPT-3G give bounds of order $10^{-10}$, and the optimistic science goals of the CORE and the CMB-S4 experiments give detection limits of order $10^{-11}$. As we cautioned before, these bounds rely on assuming extra knowledge so strong that the best $G\mu$ inference method will yield a weaker limit. To achieve a detection limit for the models described in~\cite{Copeland:2004dw,Witten:1985tq} we will need to go beyond CMB measurements, perhaps to the future promise of the information rich 21 cm intensity maps. 

\section*{Acknowledgements}
We thank Fran\c{c}ois Bouchet and Christophe Ringeval 
for generously providing us the CMB maps and the string temperature maps that we used in the convolutional neural network analysis of section~\ref{sec:CNNnoise} and the information theory bound calculation of section~\ref{sec:infoBound}.
We also thank Robert Brandenberger and Jonathan Seviers for helpful discussions. We acknowledge the support of the Fonds de recherche du Qu\'ebec -- Nature et technologies (FRQNT) Programme de recherche pour les enseignants de coll\`ege, and the support of the Natural Sciences and Engineering Research Council of Canada (NSERC) (funding reference number SAPIN-2018-00020).
Computations were made on the supercomputer Helios from Universit\'e Laval, managed by Calcul Qu\'ebec and Compute Canada. The operation of this supercomputer is funded by the Canada Foundation for Innovation (CFI), the minist\`ere de l'\'Economie, de la science et de l'innovation du Qu\'ebec (MESI) and the Fonds de recherche du Qu\'ebec -- Nature et technologies (FRQNT). 

\bibliographystyle{mnras}
\bibliography{cs_info_bound}


\bsp	
\label{lastpage}
\end{document}